\documentclass[twocolumn,preprintnumbers,pra,hyperref]{revtex4}
\usepackage{amssymb}
\usepackage{amsfonts}
\usepackage{amsmath}
\usepackage{graphicx}

\begin{document}

\title{Nonclassicality and decoherence of photon-added squeezed thermal
state in thermal environment}
\author{Li-Yun Hu$^{1,2}$\thanks{{\small E-mail: hlyun2008@126.com.}} and
Zhi-Ming Zhang$^{2}$\thanks{{\small E-mail: zmzhang@scnu.edu.cn}}}
\affiliation{$^{1}${\small College of Physics \& Communication Electronics, Jiangxi
Normal University, Nanchang 330022, China}\\
$^{2}${\small Key Laboratory of Photonic Information Technology of Guangdong
Higher Education Institutes, }\\
{\small SIPSE \& LQIT, South China Normal University, Guangzhou 510006, China%
}}

\begin{abstract}
{\small Theoretical analysis is given of nonclassicality and decoherence of
the field states generated by adding any number of photons to the squeezed
thermal state (STS). Based on the fact that the squeezed number state can be
considered as a single-variable Hermite polynomial excited state, the
compact expression of the normalization factor is derived, a Legendre
polynomial. The nonclassicality is investigated by exploring the
sub-Poissonian and negative Wigner function (WF). The results show that the
WF of single photon-added STS (PASTS) always has negative values at the
phase space center. The decoherence effect on PASTS is examined by the
analytical expression of WF. It is found that a longer threshold value of
decay time is included in single PASTS than in single-photon subtraction STS.%
}

PACS number(s): 42.50.Dv, 03.65.Wj, 03.67.Mn
\end{abstract}

\maketitle

\section{Introduction}

Generation and manipulation of non-classical light field has been a topic of
great interest in quantum optics and quantum information science \cite{1}.
Many experimental schemes have been proposed to generate nonclassical states
of optical field. Among them, subtracting photons from and/or adding photons
to quantum states have been paid much attention because these fields exhibit
an abundant of nonclassical properties and may give access to a complete
engineering of quantum states and to fundamental quantum phenomena \cite%
{2,3,4,5,6,7,8,9,10}. For example, quantum-to-classical transition has been
realized experimentally through single-photon-added coherent states of
light. These states allow one to witness the gradual change from the
spontaneous to the stimulated regimes of light emission \cite{4}. For $m$%
-photon-added coherent state in the dissipative channel, the nonclassical
properties are studied theoretically \cite{11} by deriving the analytical
expression of the Wigner function (WF), which turns out to be a
Laguerre-Gaussian function. As another example, photon addition and
subtraction experimentally have been employed to probe quantum commutation
rules by Parigi \textit{et al}. In fact, they have implemented simple
alternated sequences of photon creation (addition) and annihilation
(subtraction) on a thermal field and observed the noncommutativity of the
creation and annihilation operators \cite{6}. In addition, photon
subtraction/addition can be applied to improve entanglement between Gaussian
states \cite{12,13}, loophole-free tests of Bell's inequality \cite{14,15},
and quantum computing \cite{16}.

On the other hand, it is interesting to notice that subtracting or adding
one photon from/to pure squeezed vacuum can generate the same output state,
i.e., squeezed single-photon state \cite{17}. Actually, the photon addition
is able to generate a nonclassical state (e.g coherent and thermal states),
which is quite different from photon subtraction only from a nonclassical
state \cite{18,19,20}. In addition, the resulting states obtained by
successive photon subtractions or additions are different from each other.
For instance, successive two-photon additions [$a^{\dag2}$] and successive
two-photon subtractions [$a^{2}$] will result in the same state produced by
using subtraction-addition ($a^{\dag}a$) and addition-subtraction ($%
aa^{\dag} $), respectively. In Ref.\cite{21}, two photon-subtracted squeezed
vacuum is used to generate the squeezed superposition of coherent states
with high fidelities and large amplitudes.

In general, different non-Gaussian operators (e.g subtracting and adding
photon) will suffer different effects from the surroundings, thus it is
important to know which operator is more robust compared to the other under
an identical initial quantum state when the environment is taken into
account. Very recently, the robustness of several superposition states is
studied by using the linear entropy under a thermal environment \cite{22}.
In this paper, we shall introduce a kind of nonclassical
state---photon-addition squeezed thermal state (PASTS), generated by adding
photon to squeezed thermal state (STS) which can be considered as a
generalized Gaussian state. Then we shall investigate the nonclassical
properties and decoherence of single-mode any number PASTS under the
influence of thermal environment.

This paper is organized as follows. In Sec. II we introduce the single-mode
PASTS. By converting the PASTS to an Hermite polynomial excitation squeezed
vacuum state, we derive a compact expression for the normalization factor of
PASTS, which is an $m$-order Legendre polynomial of squeezing parameter $%
\lambda$ and mean number $n_{c}$ of thermal state, where $m$ is the number
of added photons. In Sec III, we discuss the nonclassical properties of the
PASTS in terms of sub-Poissonian statistics and the negativity of its WF. We
find the negative region of WF in phase space and there is an upper bound
value of $\lambda$ for this state to exhibit sub-Poissonian statistics which
increases as $m$ increases. Then, in Sec. IV we derive the explicitly
analytical expression of time evolution of WF of the arbitrary PASTS in the
thermal channel and discuss the loss of nonclassicality in reference of the
negativity of WF. The threshold value of decay time corresponding to the
transition of the WF from partial negative to completely positive definite
is obtained at the center of the phase space, which is independent of
parameters $\lambda$ and $n_{c}$. It shown that the WF for single PASTS
(SPASTS) has always negative value for all parameters $\lambda$ and $n_{c}$
if the decay time $\kappa t<\frac{1}{2}\ln[(2\mathcal{N}+2)/(2\mathcal{N}+1)]%
\ $(see Eq.(\ref{7.10}) below), where $\mathcal{N}$ denotes the average
thermal photon number in the environment\textbf{\ }with dissipative
coefficient\textbf{\ }$\kappa$. Comparing to the case of single
photon-subtraction STS (SPSSTS), the decoherence time of SPASTS is longer.
In this sense, the photon-addition non-Gaussian states present more robust
contrast to decoherence than photon-subtraction ones. The reason may be that
the amount of non-Gaussianity for SPASTS is larger than that for SPSSTS as
presented in Sec. V. Conclusions are involved in the last section.

\section{Photon-addition squeezed thermal state (PASTS)}

The $m$-photon-added scheme, denoted by the mapping $\rho \rightarrow
a^{\dag m}\rho a^{m},$ was first proposed by Agarwal and Tara \cite{18}.
Here, we introduce the PASTS. Theoretically, the PASTS can be obtained by
repeatedly operating the photon creation operator $a^{\dagger }$ on a STS,
so its density operator is given by%
\begin{equation}
\rho _{ad}=C_{a,m}^{-1}a^{\dag m}S_{1}^{\dagger }\rho _{th}S_{1}a^{m},
\label{2.5}
\end{equation}%
where $m$ is the added photon number (a non-negative integer), $C_{a,m}^{-1}$
is the normalization constant to be determined, and $S_{1}=\exp [\lambda
(a^{2}-a^{\dagger 2})/2]$ is the single-mode squeezing operator with $%
\lambda $ being squeezing parameter \cite{23,24}. $\rho _{th}$ is a single
field mode with frequency $\omega $ in a thermal equilibrium state
corresponding to absolute temperature $T$, whose the density operator is
\cite{25}
\begin{equation}
\rho _{th}=\sum_{n=0}^{\infty }\frac{n_{c}^{n}}{\left( n_{c}+1\right) ^{n+1}}%
\left\vert n\right\rangle \left\langle n\right\vert =\frac{1}{n_{c}}\vdots
e^{-\frac{1}{n_{c}}a^{\dag }a}\vdots ,  \label{2.1}
\end{equation}%
($\vdots $ $\vdots $ denoting antinormally ordering) which implies that the
density operator $\rho _{th}$ can be expanded as%
\begin{equation}
\rho _{th}=\frac{1}{n_{c}}\int \frac{d^{2}\alpha }{\pi }e^{-\frac{1}{n_{c}}%
\left\vert \alpha \right\vert ^{2}}\left\vert \alpha \right\rangle
\left\langle \alpha \right\vert ,  \label{2.6}
\end{equation}%
where $n_{c}=[\exp (\omega /(kT))-1]^{-1}$ being the average photon number
of the thermal state $\rho _{th}$ and $k_{B}$ being Boltzmann's constant.
Eq.(\ref{2.6}) is useful for later calculation.

\subsection{Squeezed number state as a Hermite polynomial excited state}

Recalling that the single-mode squeezed operator $S_{1}$ has its natural
expression in the coordinate representation \cite{26},
\begin{equation}
S_{1}=\frac{1}{\sqrt{\mu }}\int_{-\infty }^{\infty }dq\left\vert \frac{q}{%
\mu }\right\rangle \left\langle q\right\vert ,\mu =e^{\lambda },  \label{3.1}
\end{equation}%
where $\left\vert q\right\rangle $ is the eigenstate of $Q=(a+a^{\dag })/%
\sqrt{2}$, \ $Q\left\vert q\right\rangle =q\left\vert q\right\rangle ,$ and
\begin{equation}
\left\vert q\right\rangle =\pi ^{-1/4}\exp \left\{ -\frac{q^{2}}{2}+\sqrt{2}%
qa^{\dagger }-\frac{a^{\dagger 2}}{2}\right\} \left\vert 0\right\rangle .
\label{3.2}
\end{equation}%
Thus, using Eq.(\ref{3.2}) and the overlap relation%
\begin{equation}
\left\langle q\right\vert \left. n\right\rangle =\frac{1}{\sqrt{2^{n}n!\sqrt{%
\pi }}}e^{-q^{2}/2}H_{n}\left( q\right) ,  \label{3.3}
\end{equation}%
where $H_{n}\left( q\right) $ is the single-variable Hermite polynomial then
$S_{1}\left\vert n\right\rangle $ can be expressed as%
\begin{align}
S_{1}\left\vert n\right\rangle & =\int_{-\infty }^{\infty }\frac{dq}{\sqrt{%
2^{n}n!\mu \sqrt{\pi }}}e^{-q^{2}/2}H_{n}\left( q\right) \left\vert \frac{q}{%
\mu }\right\rangle   \notag \\
& =\frac{\text{sech}^{1/2}\lambda }{\sqrt{2^{n}n!}}\frac{\partial ^{n}}{%
\partial \tau ^{n}}\left. e^{\sqrt{2}a^{\dagger }\tau \text{sech}\lambda
+(\tau ^{2}-\frac{1}{2}a^{\dagger 2})\tanh \lambda }\left\vert
0\right\rangle \right\vert _{\tau =0}  \notag \\
& =\frac{\left( i\sqrt{\tanh \lambda }\right) ^{n}}{\sqrt{2^{n}n!}}%
H_{n}\left( \frac{a^{\dagger }\text{sech}\lambda }{i\sqrt{2\tanh \lambda }}%
\right) S_{1}\left\vert 0\right\rangle ,  \label{3.4}
\end{align}%
where we have set $\text{sech}\lambda =2\mu /(\mu ^{2}+1),$ $\tanh \lambda
=(\mu ^{2}-1)/(\mu ^{2}+1),$ and we have used $S_{1}\left\vert
0\right\rangle =\text{sech}^{1/2}\lambda \exp [-a^{\dagger 2}/2\tanh \lambda
]\left\vert 0\right\rangle $ as well as the generating function of $%
H_{n}\left( q\right) $ \cite{27}:%
\begin{equation}
H_{n}\left( q\right) =\left. \frac{\partial ^{n}}{\partial \tau ^{n}}\exp
\left( 2q\tau -\tau ^{2}\right) \right\vert _{\tau =0}.  \label{3.5}
\end{equation}%
Eq.(\ref{3.4}) indicates that the single-mode squeezed number state $%
S_{1}\left\vert n\right\rangle $ is actually a Hermite polynomial excited
squeezed vacuum state \cite{28}. Obviously, when $n=0,H_{0}\left( q\right)
=1,$ Eq.(\ref{3.4}) just reduces to single-mode squeezed vacuum. While for $%
n=1,2,$noting $H_{1}\left( q\right) =2q$ and $H_{2}\left( q\right) =4q^{2}-2,
$ Eq.(\ref{3.4}) become%
\begin{align}
S_{1}\left\vert 1\right\rangle & =a^{\dagger }\text{sech}\lambda \text{ }%
S_{1}\left\vert 0\right\rangle ,  \notag \\
S_{1}\left\vert 2\right\rangle & =\frac{1}{\sqrt{2}}\left( a^{\dagger 2}%
\text{sech}^{2}\lambda +\tanh \lambda \right) S_{1}\left\vert 0\right\rangle
,  \label{3.6}
\end{align}%
respectively. It is interesting to notice that the single photon-added
squeezed vacuum (PASV) is equal to the squeezed number state $%
S_{1}\left\vert 1\right\rangle $, and the two PASV can be considered as a
superposition of the squeezed number state $S_{1}\left\vert 2\right\rangle $
and the squeezed vacuum.

\subsection{Normalization of PASTS}

To fully describe a quantum state, its normalization is usually necessary.
Next, we shall employ the fact (\ref{3.4}) to realize our aim. First, let us
derive the normally ordering form of STS $\rho_{s}\equiv
S_{1}^{\dagger}\rho_{th}S_{1}$, which is convenient for further calculation
of normalization.

Using Eqs.(\ref{2.1}) and (\ref{3.4}), we can rewrite the STS $\rho _{s}$ as%
\begin{align}
\rho _{s}& =\sum_{n=0}^{\infty }\frac{n_{c}^{n}}{\left( n_{c}+1\right) ^{n+1}%
}S_{1}\left( -\lambda \right) \left\vert n\right\rangle \left\langle
n\right\vert S_{1}^{\dag }\left( -\lambda \right)   \notag \\
& =\frac{\text{sech}\lambda }{n_{c}+1}\sum_{n=0}^{\infty }\frac{\left(
n_{c}\tanh \lambda \right) ^{n}}{2^{n}n!\left( n_{c}+1\right) ^{n}}\colon
H_{n}\left( \frac{-a^{\dagger }\text{sech}\lambda }{\sqrt{2\tanh \lambda }}%
\right)   \notag \\
& \times \exp \left[ \frac{1}{2}\left( a^{2}+a^{\dagger 2}\right) \tanh
\lambda -a^{\dag }a\right] H_{n}\left( \frac{-a\text{sech}\lambda }{\sqrt{%
2\tanh \lambda }}\right) \colon ,  \label{5.1}
\end{align}%
where $S_{1}^{\dag }\left( -\lambda \right) =S_{1}\left( \lambda \right) $
and the vacuum projector $\left\vert 0\right\rangle \left\langle
0\right\vert =\colon \exp \left[ -a^{\dag }a\right] \colon $ is used.
Further using the two-linear generating function of Hermite polynomial \cite%
{29},%
\begin{align}
& \sum_{n=0}^{\infty }\frac{t^{n}}{2^{n}n!}H_{n}\left( x\right) H_{n}\left(
y\right)   \notag \\
& =\frac{1}{\sqrt{1-t^{2}}}\exp \left[ \frac{2txy-t^{2}\left(
x^{2}+y^{2}\right) }{1-t^{2}}\right] ,  \label{5.2}
\end{align}%
we can directly obtain the normally ordering form of STS,%
\begin{equation}
\rho _{s}=\frac{1}{\sqrt{A}}\colon \exp \left[ \frac{C}{2}\left( a^{\dagger
2}+a^{2}\right) +\left( B-1\right) a^{\dagger }a\right] \colon ,  \label{5.3}
\end{equation}%
where we have set
\begin{align}
A& =n_{c}^{2}+\left( 2n_{c}+1\right) \cosh ^{2}\lambda ,  \notag \\
B& =\frac{n_{c}}{A}\left( n_{c}+1\right) ,  \notag \\
C& =\frac{\allowbreak 2n_{c}+1}{2A}\sinh 2\lambda .  \label{5.3a}
\end{align}%
By introducing $a=(Q+iP)/\sqrt{2}$ and $a^{\dagger }=(Q-iP)/\sqrt{2}$, Eq.(%
\ref{5.3}) can be put into another form%
\begin{equation}
\rho _{s}=\frac{1}{\tau _{1}\tau _{2}}\colon \exp \left[ -\frac{Q^{2}}{2\tau
_{1}^{2}}-\frac{P^{2}}{2\tau _{2}^{2}}\right] \colon ,  \label{5.4}
\end{equation}%
where $\tau _{1}\tau _{2}=\sqrt{A},$ and
\begin{align}
2\tau _{1}^{2}& =\left( \allowbreak 2n_{c}+1\right) e^{2\lambda }+1,  \notag
\\
2\tau _{2}^{2}& =\left( \allowbreak 2n_{c}+1\right) e^{-2\lambda }+1.
\label{5.5}
\end{align}%
Eq.(\ref{5.3}) or (\ref{5.4}) is a compact expression of the STS, which is
just a Gaussian distribution within normal order for operators $Q$ and $P$
\cite{30}.

Next, we shall derive the normalization factor for PASTS. Employing Eq.(\ref%
{5.3}), the PASTS reads as
\begin{equation}
\rho _{ad}=\frac{C_{a,m}^{-1}}{\tau _{1}\tau _{2}}\colon a^{\dag m}\exp %
\left[ \frac{C}{2}\left( a^{\dagger 2}+a^{2}\right) +\left( B-1\right)
a^{\dagger }a\right] a^{m}\colon .  \label{5.6}
\end{equation}%
Thus the normalization factor $C_{a,m}$ is $\left( 1=\mathtt{tr}\rho
_{ad}\right) $
\begin{align}
C_{a,m}& =\frac{1}{\tau _{1}\tau _{2}}\int \frac{d^{2}\alpha }{\pi }%
\left\vert \alpha \right\vert ^{2m}e^{-\left( 1-B\right) \left\vert \alpha
\right\vert ^{2}+\frac{C}{2}\left( \alpha ^{\ast 2}+\alpha ^{2}\right) }
\notag \\
& =\frac{\partial ^{2m}}{\partial s^{m}\partial t^{m}}\int \frac{d^{2}\alpha
}{\pi \tau _{1}\tau _{2}}\left. e^{-\left( 1-B\right) \left\vert \alpha
\right\vert ^{2}+s\alpha ^{\ast }+t\alpha +\frac{C}{2}\left( \alpha ^{\ast
2}+\alpha ^{2}\right) }\right\vert _{s=t=0}  \notag \\
& =\frac{\partial ^{2m}}{\partial s^{m}\partial t^{m}}\exp \left[ A\left(
1-B\right) st+\frac{AC}{2}\left( s^{2}+t^{2}\right) \right] _{s=t=0},
\label{5.7}
\end{align}%
where we have used the completeness relation of coherent state, and $[\left(
1-B\right) ^{2}-C^{2}]^{-1}=\tau _{1}^{2}\tau _{2}^{2}=A$, as well as the
integration formula \cite{31}%
\begin{align}
& \int \frac{d^{2}z}{\pi }\exp \left( \zeta \left\vert z\right\vert ^{2}+\xi
z+\eta z^{\ast }+fz^{2}+gz^{\ast 2}\right)  \notag \\
& =\frac{1}{\sqrt{\zeta ^{2}-4fg}}\exp \left[ \frac{-\zeta \xi \eta +\xi
^{2}g+\eta ^{2}f}{\zeta ^{2}-4fg}\right] ,  \label{4.5}
\end{align}%
whose convergent condition is Re$\left( \zeta \pm f\pm g\right) <0,\ $Re($%
\zeta ^{2}-4fg)/(\zeta \pm f\pm g)<0.$

Recalling the newly found formula of Legendre polynomial \cite{32,33}, i.e.,%
\begin{align}
& \frac{\partial^{2m}}{\partial t^{m}\partial\tau^{m}}\left. \exp\left(
-t^{2}-\tau^{2}+\frac{2x\tau t}{\sqrt{x^{2}-1}}\right) \right\vert
_{t,\tau=0}  \notag \\
& =\frac{2^{m}m!}{\left( x^{2}-1\right) ^{m/2}}P_{m}\left( x\right) ,
\label{4.6}
\end{align}
and noticing $x^{2}-1=AC^{2},$ together with $x=\sqrt{A}\left( 1-B\right) =%
\left[ A-n_{c}\left( n_{c}+1\right) \right] /\sqrt{A}$, we have%
\begin{align}
C_{a,m} & =\frac{\left( AC\right) ^{m}}{2^{m}}\frac{\partial^{2m}}{\partial
s^{m}\partial t^{m}}\exp\left[ \frac{2}{C}\left( 1-B\right) st-s^{2}-t^{2}%
\right] _{s=t=0}  \notag \\
& =m!A^{m/2}P_{m}\left( \bar{B}/\sqrt{A}\right) ,  \label{5.8}
\end{align}
which indicates that $C_{a,m}$ is also just related to Legendre polynomial,
and
\begin{equation}
\bar{B}=n_{c}\cosh2\lambda+\cosh^{2}\lambda.  \label{5.9}
\end{equation}
It is noted that, for the case of no-photon-addition with $m=0$, $C_{a,0}=1$
as expected. Under the case of $m$-photon-addition thermal state (no
squeezing) with $\bar{B}=\allowbreak n_{c}+1$, $A=\allowbreak\left(
n_{c}+1\right) ^{2},$ and $P_{m}\left( 1\right) =1$, then $C_{a,m}=m!\left(
n_{c}+1\right) ^{m}.$ The same result as Eq.(32) can be found in Ref.\cite%
{34}.

\section{Nonclassical properties of PASTS}

In this section, we shall discuss the nonclassical properties of PASTS in
terms of sub-Poissonian statistics and the negativity of its WF.

\subsection{Sub-Poissonian nature of PASTS}

The nonclassicality of the PASTS can be analyzed by studying its
sub-Poissonian distribution. Using Eq.(\ref{5.8}) we can directly calculate:
\begin{align}
\left\langle a^{\dag}a\right\rangle & =\frac{C_{a,m+1}}{C_{a,m}}-1,
\label{2.2} \\
\left\langle a^{\dag2}a^{2}\right\rangle & =\frac{C_{a,m+2}}{C_{a,m}}-4\frac{%
C_{a,m+1}}{C_{a,m}}+2.  \label{2.3}
\end{align}
Thus the Mandel's $\mathcal{Q}$-parameter \cite{35} can be obtained by
substituting Eqs.(\ref{2.2}) into $\mathcal{Q}\equiv\left\langle a^{\dagger
2}a^{2}\right\rangle /\left\langle a^{\dag}a\right\rangle -\left\langle
a^{\dag}a\right\rangle ,$%
\begin{equation}
\mathcal{Q=}\frac{C_{a,m+2}-4C_{a,m+1}+2C_{a,m}}{C_{a,m+1}-C_{a,m}}-\frac{%
C_{a,m+1}-C_{a,m}}{C_{a,m}}.  \label{2.4}
\end{equation}
The negativity of the Mandel's $\mathcal{Q}$-parameter refers to
sub-Poissonian statistics of the state. In order to see clearly the
variation of $\mathcal{Q}$-parameter with $\lambda$ and $n_{c}$, we show the
plots of $\mathcal{Q}$-parameter in Fig.1, from which one can clearly see
that, for a given small $n_{c}$ value, $\mathcal{Q}$-parameter\ becomes
negative ($m\neq0)$ when $\lambda$ is less than a certain threshold value
which increases as $m$ increases; while for $m=0\ $or a large $n_{c}$, $%
\mathcal{Q}$ is always positive. This implies that the nonclassicality is
enhanced by adding photon to squeezed state. Here, we should emphasize that
the WF has negative region for all $\lambda$ and $n_{c},$ and thus the PASTS
is nonclassical.
\begin{figure}[tbp]
\label{Fig0} \centering
\includegraphics[width=8cm]{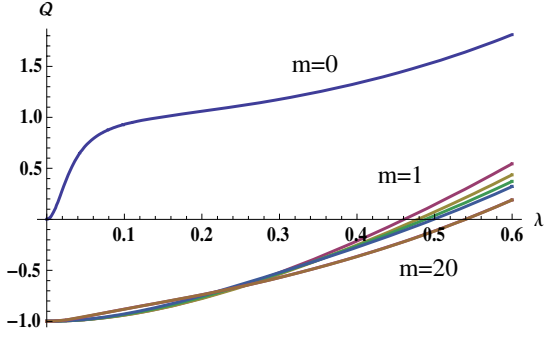}
\caption{{\protect\small (Color online) The }$\mathcal{Q}${\protect\small %
-parameter as the function of squeezing parameter}$r${\protect\small \ for
different }$m=0,1,2,3,4,19,20$ with a small $n_{c}$ value.}
\end{figure}

\subsection{Photon-number distribution (PND) of the PASTS}

The photon-number distribution (PND) is a key characteristic of every
optical field. For this purpose, we first calculate the PND of STS, then the
PND of PASTS can be\ directly obtain. The PND, i.e., the probability of
finding $n$ photons in a quantum state described by the density operator $%
\rho $, is $\mathcal{P}(n)=\left\langle n\right\vert \rho \left\vert
n\right\rangle .$ So the PND of the STS is
\begin{equation}
\mathcal{P}(n)=\left\langle n\right\vert S_{1}^{\dagger }\rho
_{th}S_{1}\left\vert n\right\rangle .  \label{4.2}
\end{equation}%
Using the fact in (\ref{3.4}) and the P-representation of $\rho _{th}$ (\ref%
{2.6}), Eq.(\ref{4.2}) can be directly written as%
\begin{align}
\mathcal{P}(n)& =\frac{\text{sech}\lambda }{2^{n}n!n_{c}}\frac{\partial ^{2n}%
}{\partial t^{n}\partial \tau ^{n}}\exp \left[ \left( t^{2}+\tau ^{2}\right)
\tanh \lambda \right]   \notag \\
& \times \int \frac{d^{2}\alpha }{\pi }\exp \left[ \sqrt{2}\left( \alpha
t+\alpha ^{\ast }\tau \right) \text{sech}\lambda -\frac{n_{c}+1}{n_{c}}%
\left\vert \alpha \right\vert ^{2}\right]   \notag \\
& \times \exp \left[ -\frac{\tanh \lambda }{2}\left( \alpha ^{2}+\alpha
^{\ast 2}\right) \right] _{\tau =t=0}  \notag \\
& =\frac{\text{sech}\lambda }{2^{n}n!\sqrt{A}}\frac{\partial ^{2n}}{\partial
t^{n}\partial \tau ^{n}}\exp \left[ 2Bt\tau +C\left( t^{2}+\tau ^{2}\right) %
\right] _{\tau =t=0}.  \label{4.3}
\end{align}%
In a similar way to deriving Eq.(\ref{5.8}), using Eq.(\ref{4.6}) we have%
\begin{equation}
\mathcal{P}(n)=\frac{D^{n/2}}{\sqrt{A}}P_{n}\left( B/\sqrt{D}\right) ,
\label{4.7}
\end{equation}%
where%
\begin{equation}
D=\frac{n_{c}^{2}-\left( 2n_{c}+1\right) \sinh ^{2}\lambda }{%
n_{c}^{2}+\left( 2n_{c}+1\right) \cosh ^{2}\lambda }.  \label{4.4}
\end{equation}%
Eq.(\ref{4.7}) shows that the PND of STS is the Legendre polynomial of $B/%
\sqrt{D}.$ In particular, when $\lambda =0,$ $A=(n_{c}+1)^{2}$ and $B/\sqrt{D%
}=1,D=n_{c}^{2}/(n_{c}+1)^{2},$ then Eq.(\ref{4.7}) becomes $\mathcal{P}%
(n)=n_{c}^{n}/(n_{c}+1)^{n},$ corresponding to the PND of thermal state \cite%
{34}. In fact, we can also check Eq.(\ref{4.7}) using the normalization
condition. Note that the Legendre polynomial can also be defined as the
coefficients in a Taylor series expansion \cite{36}
\begin{equation}
\frac{1}{\sqrt{1-2xt+t^{2}}}=\sum_{n=0}^{\infty }P_{n}\left( x\right) t^{n},
\label{4.8}
\end{equation}%
thus $\sum_{n=0}^{\infty }\mathcal{P}(n)=1/\sqrt{A(1-2B+D)}=1$ as expected.

Next, we turn to present the PND of PASTS. From Eq.(\ref{4.7}) and noting $%
a^{\dag m}\left\vert n\right\rangle =\sqrt{(m+n)!/n!}\left\vert
m+n\right\rangle $ and $a^{m}\left\vert n\right\rangle =\sqrt{n!/(n-m)!}%
\left\vert n-m\right\rangle $, it then directly follows%
\begin{align}
\mathcal{P}_{2}(n)& =C_{a,m}^{-1}\left\langle n\right\vert a^{\dag m}\rho
_{s}a^{m}\left\vert n\right\rangle  \notag \\
& =\frac{n!C_{a,m}^{-1}D^{(n-m)/2}}{(n-m)!\sqrt{A}}P_{n-m}\left( B/\sqrt{D}%
\right) .  \label{4.11}
\end{align}%
Eq.(\ref{4.11}) is the PND of PASTS, a Legendre polynomial with a condition $%
n\geqslant m$ which implies that the photon-number ($n$) involved in PASTS
is always no-less than the photon-number ($m$) operated on the STS, and
there is no photon distribution when $n<m$)\textbf{.} For some other
non-Gaussian states, such as $a^{\dag n}a^{m}\rho _{s}a^{\dag
m}a^{n},a^{m}a^{\dag n}\rho _{s}a^{n}a^{\dag m},$ and $a^{m}\rho _{s}a^{\dag
m},$ their PNDs can also be directly obtained by using Eq.(\ref{4.7}). In
Fig. 2, the PND is shown for different values $\left( \lambda ,n_{c}\right) $
and $m.$ By adding photons, we have been able to move the peak from zero
photons to nonzero photons (see blue and red bar in Fig.2). The position of
peak depends on how many photons are created and how much the state is
squeezed initially. The probability of PND becomes smaller with the
increasement of squeezing parameter (see red and green bar in Fig.2).
\begin{figure}[tbp]
\label{Fig1} \centering\includegraphics[width=8cm]{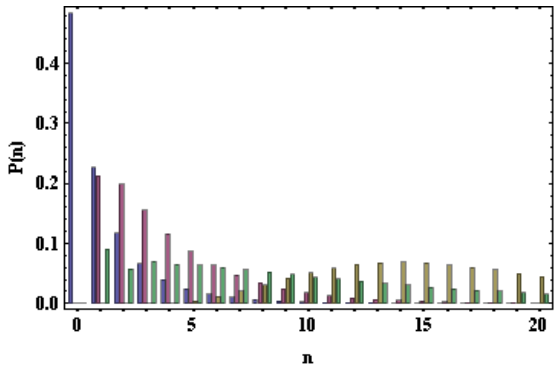}
\caption{{\protect\small (Color online) Photon-number distributions of PASTS
with \={n}=1 for }${\protect\small \protect\lambda }${\protect\small =0.3,
m=0 (blue bar); }${\protect\small \protect\lambda }${\protect\small =0.3,
m=1 (red bar), }${\protect\small \protect\lambda }${\protect\small =0.3, m=5
(yellow bar), and }${\protect\small \protect\lambda }${\protect\small =0.8,
m=1 (green bar). }}
\end{figure}

\section{Wigner function of PASTS}

Next, the normally ordering form Eq.(\ref{5.3}) is applied to deduce the WF
of PASTS. The partial negativity of WF is indeed a good indication of the
highly nonclassical character of the state. Therefore it is worth of
obtaining the WF for any states. The WF $W\left( \alpha,\alpha^{\ast}\right)
$ associated with a quantum state $\rho$ can be derived as follows \cite{23}:%
\begin{equation}
W\left( \alpha,\alpha^{\ast}\right) =e^{2\left\vert \alpha\right\vert
^{2}}\int\frac{\mathtt{d}^{2}\beta}{\pi^{2}}\left\langle -\beta\right\vert
\rho\left\vert \beta\right\rangle e^{2\left( \alpha\beta^{\ast}-\alpha^{\ast
}\beta\right) },  \label{6.1}
\end{equation}
where $\left\vert \beta\right\rangle =\exp(-\left\vert \beta\right\vert
^{2}/2+\beta a^{\dag})\left\vert 0\right\rangle $ is the coherent state.

Substituting Eq.(\ref{5.6}) into Eq.(\ref{6.1}), we can finally obtain the
WF of PASTS (see Appendix A),
\begin{equation}
W\left( \alpha,\alpha^{\ast}\right) =F_{m}\left( \alpha,\alpha^{\ast
}\right) W_{0}\left( \alpha,\alpha^{\ast}\right) ,  \label{6.2}
\end{equation}
where $W_{0}\left( \alpha,\alpha^{\ast}\right) $ is the WF of STS,
\begin{align}
W_{0}\left( \alpha,\alpha^{\ast}\right) & =\frac{1}{\pi\allowbreak\left(
\allowbreak2n_{c}+1\right) \allowbreak}\exp\left[ -\frac{2\cosh2r}{2n_{c}+1}%
\left\vert \alpha\right\vert ^{2}\right.  \notag \\
& +\left. \frac{\sinh2r}{\allowbreak2n_{c}+1}\left( \alpha^{2}+\alpha
^{\ast}{}^{2}\right) \right] ,  \label{6.3}
\end{align}
and%
\begin{align}
F_{m}\left( \alpha,\alpha^{\ast}\right) & =\frac{\left( m!\right)
^{2}C_{am}^{-1}\sinh^{m}2\lambda}{2^{2m}\left( 2n_{c}+1\right) ^{m}}  \notag
\\
& \times\sum_{l=0}^{m}\frac{\left( -1\right) ^{l}2^{2l}\left(
n_{c}+\cosh^{2}\lambda\right) ^{l}}{l!\left[ \left( m-l\right) !\right]
^{2}\sinh^{l}2\lambda}\left\vert H_{m-l}(\bar{\gamma})\right\vert ^{2},
\label{6.4}
\end{align}
where $\bar{\gamma}=[\alpha^{\ast}\sinh2\lambda-2\alpha(\cosh^{2}\lambda
+n_{c})]/\{i[\left( 2n_{c}+1\right) \sinh2\lambda]^{1/2}\}.$ Eq.(\ref{6.2})
is the analytical expression of WF for PASTS, related to single-variable
Hermite polynomials. In particular, when $m=0,$ $F_{0}\left( \alpha
,\alpha^{\ast}\right) =1,$ Eq.(\ref{6.2}) becomes $W\left( \alpha
,\alpha^{\ast}\right) =W_{0}\left( \alpha,\alpha^{\ast}\right) $; while for $%
\lambda=0$, note $C_{am}=m!\left( n_{c}+1\right) ^{m}$, $W_{0}\left(
\alpha,\alpha^{\ast}\right) =e^{-2\left\vert \alpha\right\vert
^{2}/\allowbreak\left( \allowbreak2n_{c}+1\right)
}/\allowbreak\lbrack\pi\left( \allowbreak2n_{c}+1\right) ]$ and $F_{m}\left(
\alpha,\alpha^{\ast}\right) =\left( -1\right) ^{m}/\left( 2n_{c}+1\right)
^{m}L_{m}[4\left( n_{c}+1\right) \left\vert \alpha\right\vert ^{2}/\left(
2n_{c}+1\right) ]$, Eq.(\ref{6.2}) reduces to
\begin{equation}
W\left( \alpha,\alpha^{\ast}\right) =\frac{\left( -1\right) ^{m}e^{-\frac{%
2\left\vert \alpha\right\vert ^{2}}{2\bar{n}+1}}}{\pi\allowbreak
\allowbreak\left( 2n_{c}+1\right) ^{m+1}}L_{m}\left( \frac{4\left(
n_{c}+1\right) }{2n_{c}+1}\left\vert \alpha\right\vert ^{2}\right) ,
\label{6.5}
\end{equation}
which corresponds to the WF of $m$-photon added thermal state \cite{34}, and
can be checked directly from Eq.(A3). In addition, for $m=1,$%
[single-photon-added squeezed thermal state (SPASTS)], $C_{a1}=\bar{B}$ (\ref%
{5.8}), the special WF of SPASTS is%
\begin{equation}
W_{1}\left( \alpha,\alpha^{\ast}\right) =F_{1}\left( \alpha,\alpha^{\ast
}\right) W_{0}\left( \alpha,\alpha^{\ast}\right) ,  \label{6.6}
\end{equation}
where
\begin{equation}
F_{1}\left( \alpha,\alpha^{\ast}\right) =\frac{\sinh2\lambda}{\left(
2n_{c}+1\right) \bar{B}}\left[ \left\vert \bar{\gamma}\right\vert ^{2}-\frac{%
n_{c}+\cosh^{2}\lambda}{\sinh2\lambda}\right] .  \label{6.7}
\end{equation}

Noting $\bar{B}>0$, thus from Eq.(\ref{6.7}) one can see that when the
factor $F_{1}\left( \alpha,\alpha^{\ast}\right) <0,$ the WF of SPASTS has
its negative distribution in phase space. This indicates that the WF of
SPASTS always has the negative values\ at the phase space center $\alpha=0$ (%
$\bar{\gamma}=0$)$,$ which is different from the case of
single-photon-subtracted STS with a condition $n_{c}<\sinh^{2}\lambda$ \cite%
{32}, but similar to the case of single-photon-added/subtracted squeezed
vacuum \cite{28,37}.
\begin{figure}[tbp]
\label{Fig2a} \centering\includegraphics[width=11cm]{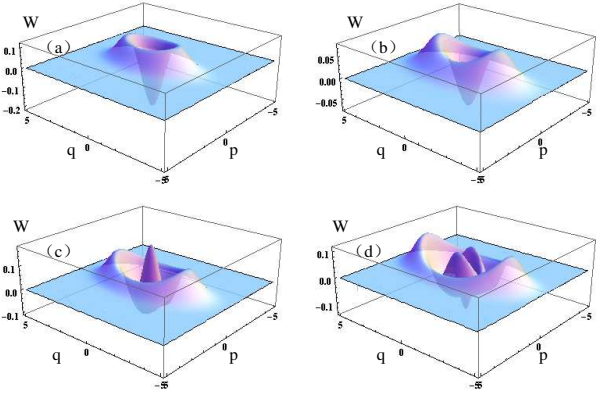}
\caption{{\protect\small (Color online) Wigner function distributions }$%
{\protect\small W}\left( \protect\alpha ,\protect\alpha ^{\ast }\right) $
{\protect\small of PASTS with }$\protect\lambda =0.3${\protect\small \ for
different }$n_{c}$ {\protect\small and }$m$ {\protect\small values (a) }$%
n_{c}=0.1,m=1;${\protect\small (b) }$n_{c}=0.5,m=1;$ {\protect\small (c) }$%
n_{c}=0.1,m=2;$ {\protect\small (d) }$n_{c}=0.1,m=3.$}
\end{figure}

Using Equations (\ref{6.2})-(\ref{6.4}) we show the plots of WF in the phase
space in Fig. 3 for the squeezing parameter ($\lambda =0.3$) with different
photon-added numbers $m$ and average number $n_{c}$ of the thermal state.
One can see clearly that there is some negative region of the WF in the
phase space which implies the nonclassicality of this state. In addition,
the squeezing effect in one of the quadratures is clear in the plots (see
Figure 3(a)), which is another evidence of the nonclassicality of this
state. The WF has its minimum value for $m=1,3$ at the center of phase space
($\alpha =0$) (see Fig. 2(a) and (d)). The case is not true for $m=2$ (see
Fig. 2(c)). For $m=2$, there are two negative regions of the WF, which
differs from the case of single PASTS. In addition, the negative region of
WF gradually decreases with the increasement of $n_{c}$, but not disappear
for $m=1$.

\section{Decoherence of PASTS in thermal environment}

In this section, we consider how this single-mode PASTS evolves at the
presence of thermal environment. In thermal channel, the evolution of the
density matrix for the $m$-PASV can be described by \cite{38}%
\begin{align}
\frac{d\rho }{dt}& =\kappa \left( \mathcal{N}+1\right) \left( 2a\rho
a^{\dagger }-a^{\dagger }a\rho -\rho a^{\dagger }a\right)  \notag \\
& +\kappa \mathcal{N}\left( 2a^{\dagger }\rho a-aa^{\dagger }\rho -\rho
aa^{\dagger }\right) ,  \label{7.1}
\end{align}%
where $\kappa $ represents the dissipative coefficient and $\mathcal{N}$
denotes the average thermal photon number of the environment. When $\mathcal{%
N}=0,$ Eq.(\ref{7.1}) reduces to the master equation describing the
photon-loss channel. The evolution formula of WF of the PASV can be derived
as follows \cite{39}
\begin{equation}
W\left( \eta ,\eta ^{\ast },t\right) =\frac{2}{\left( 2\mathcal{N}+1\right)
\mathcal{T}}\int \frac{d^{2}\alpha }{\pi }W\left( \alpha ,\alpha ^{\ast
},0\right) e^{-2\frac{\allowbreak \left\vert \eta -\alpha e^{-\kappa
t}\right\vert ^{2}}{\left( 2\mathcal{N}+1\right) \mathcal{T}}},  \label{7.2}
\end{equation}%
where $W\left( \alpha ,\alpha ^{\ast },0\right) $ is the WF of the initial
state, and $\mathcal{T}=1-e^{-2\kappa t}$. Thus, in thermal channel, the WF
at any time can be obtained by performing the integration when the initial
WF is known.

In a similar way to deriving Eq.(\ref{6.2}), substituting Eqs.(\ref{6.2})-(%
\ref{6.4}) into Eq.(\ref{7.2}) and using the generating function of
single-variable Hermite polynomials (\ref{3.5}), we finally obtain (see
Appendix B)%
\begin{equation}
W\left( \eta ,\eta ^{\ast },t\right) =F_{m}\left( \eta ,\eta ^{\ast
},t\right) W_{0}\left( \eta ,\eta ^{\ast },t\right) ,  \label{7.3}
\end{equation}%
where%
\begin{align}
W_{0}\left( \eta ,\eta ^{\ast },t\right) & =\frac{2/\left( 2n_{c}+1\right) }{%
\pi \left( 2\mathcal{N}+1\right) \mathcal{T}\sqrt{G}}  \notag \\
& \times \exp \left[ -\Delta _{1}\left\vert \eta \right\vert ^{2}+\frac{%
g_{2}g_{3}^{2}}{G}\left( \eta ^{\ast 2}+\eta ^{2}\right) \right] ,
\label{7.4}
\end{align}%
\begin{equation}
F_{m}\left( \eta ,\eta ^{\ast },t\right) =C_{am}^{-1}\sum_{l=0}^{m}\frac{%
\left( m!\right) ^{2}\chi ^{l}\Delta _{2}^{m-l}}{l!\left[ \left( m-l\right) !%
\right] ^{2}}\left\vert H_{m-l}\left( \frac{-i\omega /2}{\sqrt{\Delta _{2}}}%
\right) \right\vert ^{2},  \label{7.5}
\end{equation}%
and
\begin{align}
g_{0}& =\frac{\cosh 2\lambda }{2n_{c}+1},\text{ }g_{1}=\frac{n_{c}\mathcal{+}%
\cosh ^{2}\lambda }{2n_{c}+1},  \notag \\
g_{2}& =\frac{\sinh 2\lambda }{2n_{c}+1},\text{ }g_{3}=\frac{2e^{-\kappa t}}{%
\left( 2\mathcal{N}+1\right) \mathcal{T}},  \label{7.6}
\end{align}%
as well as
\begin{align}
G& =\left( 2g_{0}+g_{3}\allowbreak e^{-\kappa t}\right) ^{2}-4g_{2}^{2},
\notag \\
\Delta _{1}& =g_{3}e^{\kappa t}\allowbreak -\frac{g_{3}^{2}}{G}\left(
2g_{0}+g_{3}\allowbreak e^{-\kappa t}\right) ,  \notag \\
\Delta _{2}& =\frac{g_{2}}{G}\left( g_{3}e^{-\kappa t}/2-1\right) ^{2},
\label{7.8} \\
\omega & =\frac{2g_{3}}{g_{3}e^{-\kappa t}-2}\left( 2\Delta _{2}\eta ^{\ast
}+\chi \eta \right) ,  \notag \\
\chi & =\frac{2-g_{3}e^{-\kappa t}}{G}\allowbreak \left(
g_{0}+g_{1}g_{3}e^{-\kappa t}+\frac{1}{\left( 2n_{c}+1\right) ^{2}}\right) .
\notag
\end{align}%
Equation (\ref{7.3}) is just the analytical expression of WF for PASTS in
the thermal channel. It is obvious that the WF loses its Gaussian property
due to the presence of single-variable Hermite polynomials. It is
interesting to notice that $W_{0}\left( \eta ,\eta ^{\ast },t\right) $ is
actually the WF of squeezed thermal state in thermal channel corresponding
to the case without photon addition ($m=0$), $F_{0}\left( \eta ,\eta ^{\ast
},t\right) =1$; while $F_{m}\left( \eta ,\eta ^{\ast },t\right) $ is just
the non-Gaussian contribution from photon-addition. The partial negativity
of WF is fully determined by that of $F_{m}\left( \eta ,\eta ^{\ast
},t\right) $.

In particular, at the initial time $\left( t=0\right) $, noting $\left( 2%
\mathcal{N}+1\right) \mathcal{T}\sqrt{G}\rightarrow2$, $g_{3}^{2}/G%
\rightarrow1,$ and $\Delta_{1}\rightarrow2g_{0}$, $\Delta_{2}\rightarrow
\sinh2\lambda/[4(2n_{c}+1)],$ $\chi\rightarrow-(\cosh^{2}\lambda
+n_{c})/(2n_{c}+1),$ as well as $\omega/(2i\sqrt{\Delta_{2}})\rightarrow
\bar{\gamma}=[\eta^{\ast}\sinh2\lambda-2\eta(\cosh^{2}\lambda+n_{c})]/\{i[%
\left( 2n_{c}+1\right) \sinh2\lambda]^{1/2}\},$ Eqs.(\ref{7.4}) and (\ref%
{7.5}) just do reduce to Eqs.(\ref{6.3}) and (\ref{6.4}), respectively,
i.e., the WF of the PASTS. On the other hand, when $\kappa t\rightarrow
\infty,$ noticing that $\mathcal{T}\rightarrow1,G\rightarrow4/\allowbreak
\left( 2n_{c}+1\right) ^{2},\Delta_{1}\rightarrow2/\left( 2\mathcal{N}%
+1\right) ,\omega/(2i\sqrt{\Delta_{2}})\rightarrow0,\Delta_{2}\rightarrow
\frac{1}{4}\left( 2n_{c}+1\right) \sinh2\lambda,$ and $\chi\rightarrow
n_{c}\cosh2\lambda+\cosh^{2}\lambda,$ as well as $H_{m}\left( 0\right)
=\left( -1\right) ^{j}\frac{m!}{j!}\delta_{m,2j},$ then Eq.(\ref{7.3})
becomes $\allowbreak W\left( \eta,\eta^{\ast},\infty\right) =1/[\pi\left( 2%
\mathcal{N}+1\right) ]\exp[-2\left\vert \eta\right\vert ^{2}/(2\allowbreak
\mathcal{N}+1)],$ a Gaussian distribution, which is independent of
photon-addition number $m$ and corresponds to the WF of thermal state with
mean thermal photon number $\mathcal{N}$. This indicates that the system
state reduces to a thermal state with mean photon number $\mathcal{N}$ after
an enough long time interaction with the environment.

In addition, for the case of $m=1$, corresponding to the case of SPASTS, Eq.
(\ref{7.3}) just becomes%
\begin{equation}
W_{1}\left( \eta ,\eta ^{\ast },t\right) =C_{a1}^{-1}W_{0}\left( \eta ,\eta
^{\ast },t\right) \left( \left\vert \omega \right\vert ^{2}+\chi \right) .
\label{7.9}
\end{equation}%
It is obvious that when $F_{1}\left( \eta ,\eta ^{\ast },t\right) <0,$ the
WF of SPASTS in thermal channel has its negative distribution in phase
space. At the center of phase space $\eta =\eta ^{\ast }=0,$ the WF of
SPASTS always has the negative values when $\chi <0$, i.e., ($%
2-g_{3}e^{-\kappa t})/(2g_{0}+g_{3}\allowbreak e^{-\kappa t}-2g_{2})<0$
(note $2g_{0}+g_{3}\allowbreak e^{-\kappa t}-2g_{2}>0$) leading to the
following condition:%
\begin{equation}
\kappa t<\kappa t_{c}=\frac{1}{2}\ln \frac{2\mathcal{N}+2}{2\mathcal{N}+1},
\label{7.10}
\end{equation}%
which is independent of the squeezing parameter $\lambda $ and the average
photon number $n_{c}$ of thermal state, there always exist negative region
for WF in phase space and the WF of PASTS is always positive in the whole
phase space when $\kappa t\ $exceeds the threshold value $\kappa t_{c}$. Due
to this and from Eq. (\ref{7.10}), we can see how the thermal noise shortens
the threshold value of the decay time. Comparing to the time threshold value
of SPSSTS \cite{32} with the identical squeezed thermal state to that of
SPASTS,%
\begin{equation}
\kappa t_{cs}=\frac{1}{2}\ln \left[ 1-\frac{2n_{c}+1}{2\mathcal{N}+1}\frac{%
n_{c}-\sinh ^{2}\lambda }{n_{c}\cosh 2\lambda +\sinh ^{2}\lambda }\right] ,
\label{7.11}
\end{equation}%
one can find a difference of $e^{2\kappa t_{c}}-e^{2\kappa t_{cs}}:$
\begin{equation}
e^{2\kappa t_{c}}-e^{2\kappa t_{cs}}=\allowbreak \frac{2n_{c}\left(
n_{c}+1\right) }{\left( 2N+1\right) \left( n_{c}\cosh 2\lambda +\sinh
^{2}\lambda \right) },  \label{7.12}
\end{equation}%
which implies that the decoherence time of SPASTS is longer than that of
SPSSTS. In this sense, the photon-addition Gaussian states present more
robust contrast to decoherence than photon-subtraction ones.

\begin{figure}[tbp]
\label{Fig3} \centering\includegraphics[width=10cm]{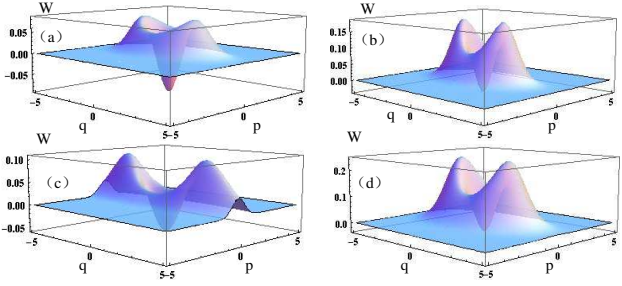}
\caption{{\protect\small (Color online) Wigner function distributions }$%
{\protect\small W}\left( \protect\alpha ,\protect\alpha ^{\ast }\right) $
{\protect\small of PASTS with }$m=1,${\protect\small \ }$n_{c}=0.3$
{\protect\small \ for different }$\mathcal{N},$ \ $\protect\lambda $
{\protect\small and }$\protect\kappa t${\protect\small \ values (a) }$%
\mathcal{N}=0.2,\protect\lambda =0.3,\protect\kappa t=0.05;${\protect\small %
(b) }$\mathcal{N}=0.2,\protect\lambda =0.3,\protect\kappa t=0.2;$
{\protect\small (c) }$\mathcal{N}=0.2,,\protect\lambda =0.8,\protect\kappa %
t=0.05;$ {\protect\small (d) }$\mathcal{N}=2,\protect\lambda =0.3,\protect%
\kappa t=0.05.$}
\end{figure}

In Fig.3, the WFs of PASTS with $m=1$ and $n_{c}=0.3$ are depicted in phase
space for several different $\mathcal{N},$ $\lambda $ and $\kappa t$\
values. It is easy to see that the negative region of WF gradually
diminishes as the time $\kappa t$ increases (see Fig.3 (a) and (b)). In
addition, the partial negativity of WF decreases gradually as $\mathcal{N}$
(or $\lambda $) increases for a given time (see Fig.3 (c) and (d)). The
squeezing effect in one of the quadrature is shown in Fig.4(c). For the case
of large squeezing value $\lambda $ and small $n_{c}$ and $\mathcal{N}$
values, the single-PASTS becomes similar to a Schodinger cat state. The WF
becomes Gaussian with the time evolution.

\section{Non-Gaussianity measure for PASTS}

As well known, non-Gaussian operators (such as photon-adding/subtracting)
can improve the nonclassicality and entanglement between Gaussian states
\cite{12,13}. One reason of such an enhancement is their amount of
non-Gaussianity \cite{40,41}. Recently, an experimentally accessible
criterion has been proposed to measure the degree based on the conditional
entropy of the state with a Gaussian reference \cite{42}. Therefore, it is
of interest to evaluate the degree of the resulting non-Gaussianity and
assess this operation as a resource to obtain non-Gaussian states starting
from Gaussian ones. Noting that the STS can be considered as a generalized
Gaussian state, thus the fidelity between PASTS and STS may be seen as a
non-Gaussianity measure. For this purpose, we define the fidelity by \cite%
{32}%
\begin{equation}
\mathcal{F}=\mathtt{tr}\left( \rho _{s}\rho \right) /\mathtt{tr}\left( \rho
_{s}^{2}\right) ,  \label{8.1}
\end{equation}%
where\ $\rho _{s}$ and $\rho $ are the STS (a generalized Gaussian state)
and the PASTS, respectively.

Noticing $\mathtt{tr}\left( \rho _{s}^{2}\right) =1/(2\bar{n}_{c}+1),$ and
using the formula (C1), we finally obtain (see Appendix C)%
\begin{equation}
\mathcal{F}=\frac{m!}{C_{a,m}}K_{2}^{m/2}P_{m}\left( \frac{K_{1}}{\sqrt{K_{2}%
}}\right) =\left( \frac{K_{2}}{A}\right) ^{m/2}\frac{P_{m}\left( K_{1}/\sqrt{%
K_{2}}\right) }{P_{m}\left( \bar{B}/\sqrt{A}\right) },  \label{8.2}
\end{equation}%
where
\begin{equation}
K_{1}=\frac{n_{c}\left( n_{c}+1\right) }{2n_{c}+1}\cosh 2\lambda ,K_{2}=%
\frac{n_{c}^{2}\left( n_{c}+1\right) ^{2}}{\left( 2n_{c}+1\right) ^{2}}%
\allowbreak -\frac{\sinh ^{2}2\lambda }{4}.  \label{8.3}
\end{equation}%
Eq.(\ref{8.2}) is just the analytical expression for the fidelity between
PASTS and STS. It is obvious that when $m=0$ (without photon-addition), $%
\mathcal{F}=1$. Comparing to the fidelity $\mathcal{F}_{s}$ between PSSTS
and STS (59) in Ref.\cite{32}, one can clearly see that%
\begin{equation}
\frac{\mathcal{F}}{\mathcal{F}_{s}}=\left( \frac{Z}{A}\right) ^{m/2}\frac{%
P_{m}\left( H/\sqrt{Z}\right) }{P_{m}\left( \bar{B}/\sqrt{A}\right) }=\frac{%
C_{s,m}}{C_{a,m}},  \label{8.4}
\end{equation}%
where $Z=n_{c}^{2}-\left( 2n_{c}+1\right) \sinh ^{2}\lambda ,$ $H=n_{c}\cosh
2\lambda +\sinh ^{2}\lambda .$ Eq.(\ref{8.4}) implies that the ratio to
fidelities is just that to the normalization factors. In particular, for $%
m=1 $ (the case of SPASTS), Eq.(\ref{8.2}) reduces to%
\begin{equation}
\frac{\mathcal{F}}{\mathcal{F}_{s}}\mathfrak{=}\frac{n_{c}\cosh 2\lambda
+\sinh ^{2}\lambda }{n_{c}\cosh 2\lambda +\cosh ^{2}\lambda }<1,  \label{8.5}
\end{equation}%
from which one can see that $\mathcal{F}\mathfrak{<}\mathcal{F}_{s},$ i.e.,
the amount of non-Gaussianity for SPASTS is larger than that for SPSSTS.

This point is made clear in Fig.5, in which the fidelity $\mathcal{F}$
between PASTS and STS as the function of squeezing parameter $\lambda $\ for
different photon-addition number $m.$ As a comparision, the fidelity $%
\mathcal{F}_{s}$ between PSSTS and STS is also shown in Fig.5, from which
one can see that the fidelity decreases as the increment of
photon-addition/subtraction number $m,$ as expected. The fidelity $\mathcal{F%
}$ increases monotonously with the augment of the squeezing parameter $%
\lambda $. However, the case is not true for the fidelity $\mathcal{F}_{s}.$
For a given $m$ value, the fidelity $\mathcal{F}$ is always smaller than the
fidelity $\mathcal{F}_{s}$ within the region shown in Fig.5. In this sense,
the amount of non-Gaussianity for PASTS is larger than that for PSSTS.

\begin{figure}[tbp]
\label{Fig5} \centering\includegraphics[width=8cm]{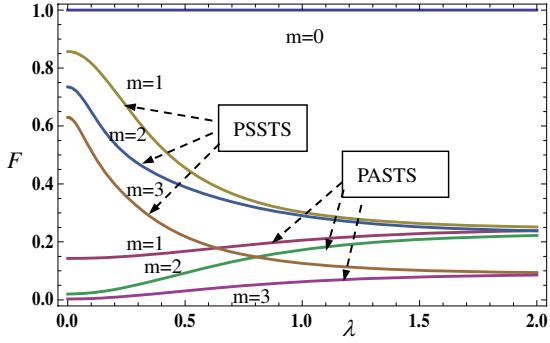}
\caption{{\protect\small (Color online) The fidelity }$\mathcal{F}$
{\protect\small \ between PASTS (PSSTS) and STS as the function of squeezing
parameter }$\protect\lambda ${\protect\small \ for different photon-addition
number }$m=0,1,2,3(n_{c}=0.2).$}
\end{figure}

\section{Conclusions}

In this paper, we investigate the nonclassical properties and decoherence of
single-mode PASTS when evolving under a thermal environment. Based on the
fact that squeezed number can be considered as an Hermite polynomial
excitation squeezed vacuum, the normally ordering form of PASTS is directly
obtained, from which one can expediently calculate some quasi-distributions,
such as Q-, P- and Wigner function; And the normalization factor of PASTS is
analytically derived, which is just proved to be an $m$-order Legendre
polynomial of the squeezing parameter $r$ and average photon number $n_{c}$
of the thermal state, a remarkable result. Furthermore, for any photon-added
number $m$-PASTS, the explicit expression of WF is derived, which considered
as a product of the WF of STS in thermal channel and a non-Gaussian
distribution resulting from photon-addition. It is shown that the WF of
SPASTS always has the negative values\ at the phase space center, which is
different from the case of SPSSTS with a condition $n_{c}<\sinh ^{2}\lambda $%
. Then the effects of decoherence to the nonclassicality of PASTS in the
thermal channel is also demonstrated according to the compact expression for
the WF. The threshold value of the decay time corresponding to the
transition of the WF from partial negative to completely positive definite
is obtained for SPASTS at the center of phase space. It is found that the WF
has always negative value for all parameters $r,n_{c}$ if the decay time $%
\kappa t<\kappa t_{c}=\frac{1}{2}\ln \frac{2\mathcal{N}+2}{2\mathcal{N}+1}$,
a larger value than that of SPSSTS.

A comparison between the nonclassicality and decoherence of PASTS and PSSTS
shows that the photon-addition non-Gaussian states present more robust
contrast to decoherence than photon-subtraction ones, which may be due to
the amount of non-Gaussianity for SPASTS is larger that that for SPSSTS. On
the other hand, in the limit of vanishing squeezing and $n_{c}=0$, the PASTS
reduces to a single-mode Fock state, remaining non-Gaussian, while the PSSTS
becomes Gaussian, as it reduces to the single mode vacuum. Entanglement
evaluation investigation for photon-subtracted/added two-mode squeezed
thermal state is a future problem.

\textbf{Acknowledgments: } This work was supported by the National Natural
Science Foundation of China (Grant Nos. 11047133, 60978009 ), the Major
Research Plan of the National Natural Science Foundation of China (Grant No.
91121023 ), and the \textquotedblleft 973\textquotedblright\ Project (Grant
No. 2011CBA00200), as well as the Natural Science Foundation of Jiangxi
Province of China (No. 2010GQW0027).

\bigskip

\textbf{Appendix A:} \textbf{Derivation of WF (\ref{6.2}) for PASTS}

Substituting Eq.(\ref{5.6}) into Eq.(\ref{6.1}) and using the integration
formula (\ref{4.5}), we have
\begin{align}
W\left( \alpha ,\alpha ^{\ast }\right) & =\frac{\left( -1\right)
^{m}C_{am}^{-1}}{\tau _{1}\tau _{2}}e^{2\left\vert \alpha \right\vert
^{2}}\int \frac{\mathtt{d}^{2}\beta }{\pi ^{2}}\left\vert \beta \right\vert
^{2m}\exp \left[ -\left( 1+B\right) \left\vert \beta \right\vert ^{2}\right.
\notag \\
& +\left. 2\left( \alpha \beta ^{\ast }-\alpha ^{\ast }\beta \right) +\frac{C%
}{2}\left( \beta ^{\ast 2}+\beta ^{2}\right) \right]  \notag \\
& =\frac{C_{am}^{-1}}{\tau _{1}\tau _{2}}e^{2\left\vert \alpha \right\vert
^{2}}\frac{\partial ^{2m}}{\partial s^{m}\partial t^{m}}\int \frac{\mathtt{d}%
^{2}\beta }{\pi ^{2}}\exp \left[ -\left( 1+B\right) \left\vert \beta
\right\vert ^{2}\right.  \notag \\
& +\left. \left( 2\alpha +s\right) \beta ^{\ast }-\left( 2\alpha ^{\ast
}+t\right) \beta +\frac{C}{2}\left( \beta ^{\ast 2}+\beta ^{2}\right) \right]
_{s=t=0}  \notag \\
& =W_{0}\left( \alpha ,\alpha ^{\ast }\right) F_{m}\left( \alpha ,\alpha
^{\ast }\right) ,  \tag{A1}
\end{align}%
where we have set
\begin{align}
W_{0}\left( \alpha ,\alpha ^{\ast }\right) & =\frac{\sqrt{A_{1}}}{\pi \tau
_{1}\tau _{2}}\exp \left[ A_{2}\left( \alpha ^{2}+\alpha ^{\ast 2}\right)
-2A_{3}\left\vert \alpha \right\vert ^{2}\right] ,  \tag{A2} \\
F_{m}\left( \alpha ,\alpha ^{\ast }\right) & =C_{am}^{-1}\frac{\partial ^{2m}%
}{\partial s^{m}\partial t^{m}}\exp \left[ \frac{A_{2}}{4}\left(
s^{2}+t^{2}\right) -\frac{A_{4}}{2}st\right.  \notag \\
& +\left. \allowbreak \left( A_{2}\alpha ^{\ast }-A_{4}\alpha \right)
t+\left( \allowbreak A_{2}\alpha -A_{4}\alpha ^{\ast }\right) s\right]
_{s=t=0},  \tag{A3}
\end{align}%
and
\begin{align}
A_{1}& =\frac{1}{\left( 1+B\right) ^{2}-C^{2}}=\frac{A}{\left(
2n_{c}+1\right) ^{2}},  \notag \\
A_{2}& =\frac{2C}{\left( 1+B\right) ^{2}-C^{2}}=\frac{\sinh 2\lambda }{%
2n_{c}+1},  \notag \\
A_{3}& =\frac{2\left( B+1\right) }{\left( 1+B\right) ^{2}-C^{2}}-1=\frac{%
\cosh 2\lambda }{2n_{c}+1},  \notag \\
A_{4}& =\frac{2\left( B+1\right) }{\left( 1+B\right) ^{2}-C^{2}}=A_{3}+1=2%
\frac{n_{c}\mathcal{+}\cosh ^{2}\lambda }{2n_{c}+1}.  \tag{A4}
\end{align}%
Substituting Eq.(A3) into Eq.(A2) yields Eq.(\ref{6.3}), i.e., the WF of
squeezed thermal state.

Further expanding the exponential term $st$ included in (A3) into sum
series, and using the generating function of single-variable Hermite
polynomials \cite{27},
\begin{equation}
H_{n}(x)=\left. \frac{\partial^{n}}{\partial t^{n}}\exp\left(
2xt-t^{2}\right) \right\vert _{t=0},  \tag{A5}
\end{equation}
which leads to%
\begin{align}
& \left. \frac{\partial^{n}}{\partial t^{n}}\exp\left( At+Bt^{2}\right)
\right\vert _{t=0}  \notag \\
& =\left( i\sqrt{B}\right) ^{n}H_{n}\left[ A/(2i\sqrt{B})\right]  \notag \\
& =\left( -i\sqrt{B}\right) ^{n}H_{n}\left[ A/(-2i\sqrt{B})\right] ,
\tag{A6}
\end{align}
thus we can see%
\begin{align}
F_{m}\left( \alpha,\alpha^{\ast}\right) & =C_{am}^{-1}\sum_{l=0}^{\infty }%
\frac{\left( -A_{4}\right) ^{l}}{2^{l}l!}\frac{\partial^{2m}}{\partial
s^{m}\partial t^{m}}s^{l}t^{l}  \notag \\
& \times\exp\left[ \frac{A_{2}}{4}\left( s^{2}+t^{2}\right) +\gamma
t+\gamma^{\ast}s\right] _{s=t=0}  \notag \\
& =C_{am}^{-1}\sum_{l=0}^{\infty}\frac{\left( -A_{4}\right) ^{l}}{2^{l}l!}%
\frac{\partial^{2l}}{\partial\gamma^{l}\partial\gamma^{\ast l}}\frac{%
\partial^{2m}}{\partial s^{m}\partial t^{m}}  \notag \\
& \times\exp\left[ \frac{A_{2}}{4}\left( s^{2}+t^{2}\right) +\gamma
t+\gamma^{\ast}s\right] _{s=t=0}  \notag \\
& =\frac{A_{2}^{m}}{2^{2m}}C_{am}^{-1}\sum_{l=0}^{\infty}\frac{\left(
-A_{4}\right) ^{l}}{2^{l}l!}\frac{\partial^{2l}}{\partial\gamma^{l}\partial%
\gamma^{\ast l}}\left\vert H_{m}\left( \bar{\gamma}\right) \right\vert ^{2},
\tag{A7}
\end{align}
where $\gamma=A_{2}\alpha^{\ast}-A_{4}\alpha,$ and $\bar{\gamma}=\gamma/(i%
\sqrt{A_{2}}),$ i.e.,
\begin{equation}
\bar{\gamma}=\frac{\alpha^{\ast}\sinh2\lambda-2\alpha\left(
\cosh^{2}\lambda+n_{c}\right) }{i\sqrt{\left( 2n_{c}+1\right) \sinh2\lambda}}%
,  \tag{A8}
\end{equation}
Then using the recurrence relation of $H_{n}(x),$
\begin{equation}
\frac{\mathtt{d}}{\mathtt{d}x^{l}}H_{n}(x)=\frac{2^{l}n!}{\left( n-l\right) !%
}H_{n-l}(x),  \tag{A9}
\end{equation}
Eq.(A7) becomes%
\begin{align}
F_{m}\left( \alpha,\alpha^{\ast}\right) & =\frac{A_{2}^{m}}{2^{2m}}%
C_{am}^{-1}\sum_{l=0}^{\infty}\frac{\left( -A_{4}/A_{2}\right) ^{l}}{2^{l}l!}
\notag \\
& \times\frac{\partial^{2l}}{\partial\bar{\gamma}^{l}}H_{m}\left( \bar{\gamma%
}\right) \frac{\partial^{2l}}{\partial\bar{\gamma}^{\ast l}}H_{m}\left( \bar{%
\gamma}^{\ast}\right)  \notag \\
& =\frac{A_{2}^{m}}{2^{2m}}C_{am}^{-1}\sum_{l=0}^{m}\frac{\left( m!\right)
^{2}\left( -2A_{4}/A_{2}\right) ^{l}}{l!\left[ \left( m-l\right) !\right]
^{2}}\left\vert H_{m-l}(\bar{\gamma})\right\vert ^{2}.  \tag{A10}
\end{align}
Substituting Eq.(A4) into Eq.(A10) yields Eq.(\ref{6.4}). Thus we complete
the derivation of WF Eq.(\textbf{\ref{6.2}}) by combing Eqs. (A2) \ and
(A10).

\textbf{Appendix B:} \textbf{Derivation of WF (\ref{7.3}) for PASTS in
thermal channel}

Substituting Eqs.(\ref{6.2})-(\ref{6.4}) into Eq.(\ref{7.2}), we have%
\begin{align}
W\left( \eta,\eta^{\ast},t\right) & =\frac{C_{am}^{-1}g_{3}e^{\kappa t}}{%
\pi\left( 2n_{c}+1\right) }e^{-g_{3}e^{\kappa t}\allowbreak\left\vert
\eta\right\vert ^{2}}\frac{\partial^{2m}}{\partial s^{m}\partial\tau^{m}}
\notag \\
& \times\exp\left[ \frac{g_{2}}{4}\left( s^{2}+\tau^{2}\right) -g_{1}s\tau%
\right]  \notag \\
& \times\int\frac{d^{2}\alpha}{\pi}\exp\left[ -\left(
2g_{0}+g_{3}\allowbreak e^{-\kappa t}\right) \left\vert \alpha\right\vert
^{2}\right.  \notag \\
& +\left( g_{3}\eta^{\ast}+g_{2}s-2g_{1}\tau\right) \alpha  \notag \\
& +\left. \allowbreak\left( g_{3}\eta+g_{2}\tau-2g_{1}s\right)
\alpha^{\ast}+g_{2}\left( \alpha^{2}+\alpha^{\ast2}\right) \right]
_{s=\tau=0},  \tag{B1}
\end{align}
where we have set%
\begin{align}
g_{0} & =A_{3}=\frac{\cosh2\lambda}{2n_{c}+1},\text{ }g_{1}=\frac{A_{4}}{2}=%
\frac{n_{c}\mathcal{+}\cosh^{2}\lambda}{2n_{c}+1},  \notag \\
g_{2} & =A_{2}=\frac{\sinh2\lambda}{2n_{c}+1},\text{ }g_{3}=\frac {%
2e^{-\kappa t}}{\left( 2\mathcal{N}+1\right) \mathcal{T}}.  \tag{B2}
\end{align}
Further using the integration (\ref{4.5}), Eq.(B1) can be put into the form%
\begin{equation}
W\left( \eta,\eta^{\ast},t\right) =F_{m}\left( \eta,\eta^{\ast},t\right)
W_{0}\left( \eta,\eta^{\ast},t\right) ,  \tag{B3}
\end{equation}
where $W_{0}\left( \eta,\eta^{\ast},t\right) $ is defined in Eq.(\ref{7.4}),
and
\begin{align}
F_{m}\left( \eta,\eta^{\ast},t\right) & =C_{am}^{-1}\frac{\partial^{2m}}{%
\partial s^{m}\partial\tau^{m}}\exp\left[ \Delta_{2}\left( s^{2}+\tau
^{2}\right) \right.  \notag \\
& +\left. \omega\tau+\omega^{\ast}s+\chi s\tau\right] _{s=\tau=0},  \tag{B4}
\end{align}
here $\left( \Delta_{2},\omega,\chi\right) $ are defined in Eq. (\ref{7.8}).
In a similar way to deriving Eq. (\ref{6.2}), we can further insert Eq. (B4)
into Eq. (\ref{7.5}).

\textbf{Appendix C: Derivation of fidelity (\ref{8.2}) between PASTS and STS}

The fidelity ($\mathtt{tr}\left( \rho_{s}\rho\right) $) can be calculated as
the overlap between the two WFs:%
\begin{equation}
\mathtt{tr}\left( \rho_{s}\rho\right) =4\pi\int d^{2}\alpha W_{0}\left(
\alpha,\alpha^{\ast}\right) W_{\rho}\left( \alpha,\alpha^{\ast}\right) ,
\tag{C1}
\end{equation}
where $W_{0}\left( \alpha,\alpha^{\ast}\right) $ is the WF of squeezed
thermal state $\rho_{s}$. Using Eq.(\ref{6.2}) we may express Eq.(C1) as%
\begin{equation}
\mathtt{tr}\left( \rho_{s}\rho\right) =4\pi\int F_{m}\left( \alpha
,\alpha^{\ast}\right) W_{0}^{2}\left( \alpha,\alpha^{\ast}\right)
d^{2}\alpha.  \tag{C2}
\end{equation}
Then employing Eqs.(\ref{6.2}) and (A2),(A3) as well as the integration
formula (\ref{4.5}), we can put Eq.(C2) into the following form:%
\begin{align}
\mathtt{tr}\left( \rho_{s}\rho\right) & =\frac{4C_{am}^{-1}}{\left(
2n_{c}+1\right) ^{2}}\frac{\partial^{2m}}{\partial s^{m}\partial\tau^{m}}\exp%
\left[ \frac{g_{2}}{4}\left( s^{2}+\tau^{2}\right) -g_{1}s\tau\right]  \notag
\\
& \int\frac{d^{2}\alpha}{\pi}\exp\left[ -4g_{0}\left\vert \alpha\right\vert
^{2}+2g_{2}\left( \alpha^{2}+\alpha^{\ast2}\right) \right]  \notag \\
& +\left. \left( \allowbreak g_{2}s-2g_{1}\tau\right) \alpha
+\allowbreak\left( g_{2}\tau-2g_{1}s\right) \alpha^{\ast}\right] _{s=\tau=0}
\notag \\
& =\frac{C_{am}^{-1}}{2n_{c}+1}\frac{\partial^{2m}}{\partial
s^{m}\partial\tau^{m}}\exp\left[ K_{1}s\tau+K_{0}\left(
s^{2}+\tau^{2}\right) \right] _{s=\tau=0},  \tag{C3}
\end{align}
where $K_{1}$ is defined in Eq.(\ref{8.3}), and
\begin{equation}
K_{0}=\frac{2n_{c}^{2}+2n_{c}+1}{4\left( 2n_{c}+1\right) }\sinh 2\lambda.
\tag{C4}
\end{equation}
Similarly to deriving Eq.(\ref{5.8}), we have%
\begin{align}
& \left. \frac{\partial^{2m}}{\partial s^{m}\partial\tau^{m}}\exp\left[
K_{0}\left( k^{2}+t^{2}\right) +\allowbreak K_{1}kt\right] \right\vert
_{k=t=0}  \notag \\
& =m!K_{2}^{m/2}P_{m}\left( K_{1}/\sqrt{K_{2}}\right) ,  \tag{C5}
\end{align}
and $K_{2}\equiv K_{1}^{2}-4K_{0}^{2}$ given in Eq.(\ref{8.3}), which leads
to Eq.(\ref{8.2}).

\end{document}